\def\year{2021}\relax
\documentclass[letterpaper]{article} 
\usepackage{aaai21}  
\usepackage{times}  
\usepackage{helvet} 
\usepackage{courier}  
\usepackage[hyphens]{url}  
\usepackage{graphicx} 
\usepackage[dvipsnames]{xcolor}
\usepackage{natbib}  
\usepackage{caption} 
\usepackage[switch]{lineno}  %

\title{In the Eyes of the Beholder: Analyzing Social Media Use of Neutral and Controversial Terms for COVID-19}
\author{
    Long Chen\textsuperscript{\rm 1}, Hanjia Lyu\textsuperscript{\rm 2}, Tongyu Yang\textsuperscript{\rm 1}, Yu Wang\textsuperscript{\rm 3}, Jiebo Luo\textsuperscript{\rm 1}\\
}
\affiliations{

    \textsuperscript{\rm 1}Department of Computer Science, University of Rochester\\
    \textsuperscript{\rm 2}Goergen Institute for Data Science, University of Rochester\\
    \textsuperscript{\rm 3}Department of Political Science, University of Rochester\\

    \{lchen62, tyang20\}@u.rochester.edu, \{hlyu5, ywang176\}@ur.rochester.edu, jluo@cs.rochester.edu

}

\begin{document}

\maketitle

\begin{abstract}
During the COVID-19 pandemic, ``Chinese Virus'' emerged as a controversial term for coronavirus. To some, it may seem like a neutral term referring to the physical origin of the virus. To many others, however, the term is in fact attaching ethnicity to the virus. While both arguments appear reasonable, quantitative analysis of the term's real-world usage is lacking to shed light on the issues behind the controversy. In this paper, we attempt to fill this gap. To model the substantive difference of tweets with controversial terms and those with non-controversial terms, we apply topic modeling and LIWC-based sentiment analysis. To test whether ``Chinese Virus'' and ``COVID-19" are interchangeable, we formulate it as a classification task, mask out these terms, and classify them using the state-of-the-art transformer models. Our experiments consistently show that the term ``Chinese Virus'' is associated with different substantive topics and sentiment compared with `COVID-19' and that the two terms are easily distinguishable by looking at their context.
\end{abstract}

\section{Introduction}

Starting in late 2019, the COVID-19 pandemic has rapidly impacted over 200 countries, areas, and territories. As of September 4, according to the World Health Organization (WHO), 26,121,999 COVID-19 cases were confirmed worldwide, with 864,618 confirmed deaths\footnote{https://www.who.int/docs/default-source/coronaviruse/situation-reports/wou-4-september-2020-approved.pdf?sfvrsn=91215c78\_2}. This disease has tremendous impacts on people’s daily lives worldwide.

In light of the deteriorating situation in the United States, discussions of the pandemic on social media have drastically increased since March 2020. Within these discussions, an overwhelming trend is the use of controversial terms targeting Asians and, specifically, the Chinese population, insinuating that the virus originated in China. On March 16, the President of the United States, Donald Trump, posted on Twitter calling COVID-19 the “Chinese Virus”.\footnote{https://twitter.com/realdonaldtrump/status/1239685852093169664} Around March 18, media coverage of the term “Chinese Flu” also took off.\footnote{https://blog.gdeltproject.org/is-it-coronavirus-or-covid-19-or-chinese-flu-the-naming-of-a-pandemic/} Although most public figures who used the controversial terms claimed them to be non-discriminative, such terms have stimulated racism and discrimination against Asian-Americans in the US, as reported by New York Times\footnote{https://www.nytimes.com/2020/03/23/us/chinese-coronavirus-racist-attacks.html}, the Washington Post\footnote{https://www.washingtonpost.com/nation/2020/03/20/coronavirus-trump-chinese-virus/}, the Guardian\footnote{https://www.theguardian.com/world/2020/mar/24/coronavirus-us-asian-americans-racism}, and other main stream news media. A recent work was done with social media data to characterize users who used controversial or non-controversial terms associated with COVID-19 and found the associations between demographics, user-level features, political following status, and geo-location attributes with the use of controversial terms~\citep{9098075}. In this study, we choose instead to analyze from a language perspective, analyzing crawled tweets (Twitter posts) with and without controversial terms associated with COVID-19.

We design our study to answer three related research questions: a) is the use of controversial terms associated with COVID-19 conveying any emotion other than mere description of the geographical origin of the virus, and b) what is the linguistic characteristics of the tweets that contain controversial terms with COVID-19, and c) whether neutral terms and controversial terms are interchangeable from a classification point of view. To answer these questions, Latent Dirichlet Allocation (LDA)~\citep{blei2003latent} is first applied to extract the topics in controversial and non-controversial posts. Next, LIWC2015 (Linguistic Inquiry and Word Count 2015)~\citep{pennebaker2015development} is applied to analyze multi-dimensional characteristics of the posts. We then make comparisons between the topics and profiles presented in both controversial and non-controversial posts, trying to investigate any association between the use of controversial terms and the underlying mindsets. Finally, we construct textual classification models with the state-of-the-art techniques to predict the use of controversial terms associated with COVID-19 on social media.

Our contributions are summarized as follows:
\begin{itemize}
    \item We analyze from a language perspective the main differences between tweets with and without controversial terms associated with COVID-19.
    \item We extract topics and LIWC features and employ state-of-the-art  models to investigate any association between the use of controversial terms and the underlying mindsets. 
    \item We discover that the controversial term is associated with different substantive topics and sentiment.
\end{itemize}

\section{Related Work}


Our work builds on previous works on text mining using data from social media during influential events. Studies have been conducted using topic modeling, a process of identifying topics in a collection of documents. The commonly used model, LDA, provides a way to automatically detect hidden topics in a given number~\citep{blei2003latent}. Previous research has been conducted on inferring topics on social media. \citet{kim2016topic} investigated the topic coverage and sentiment dynamics on Twitter and news press regarding the issue of Ebola. \citet{chen2020social} found LDA-generated topics from e-cigarette related posts on Reddit to identify potential associations between e-cigarette uses and various self-reporting health symptoms. \citet{wang2016catching} applied negative binomial regression upon abstract topics of LDA to model the “likes” on Trump’s Twitter and infer topic preferences among followers.

A large number of studies were performed with LIWC, an API\footnote{https://liwc.wpengine.com/} for linguistic analysis of documents. \citet{tumasjan2010predicting} used LIWC to capture the political sentiment and predict elections with Twitter.
Our motivation is to combine qualitative analysis with LDA and quantitative analysis with LIWC, comparatively investigate discrepancies between the tweets that use controversial terms associated with COVID-19 and the tweets that use non-controversial terms.

Previous studies have attempted to make textual classification on social media data. \citet{mouthami2013sentiment} implemented a classification model that approximately classifies the sentiment using Bag of words in Support Vector Machine (SVM) algorithm. \citet{huang2014cyber} applied SMOTE (Synthetic Minority Oversampling TEchnique) method to defecting online cyber-bullying behavior. In addition, a number of other studies performed textual classifications for various purposes using social media data~\citep{chatzakou2015harvesting, lukasik2016hawkes, zhang2016user}.

\section{Data and Methodology}

In this section, we describe data collection, pre-processing and methods to analyze data.

\begin{figure*}[htbp]
    \centering
    \includegraphics[width=\linewidth]{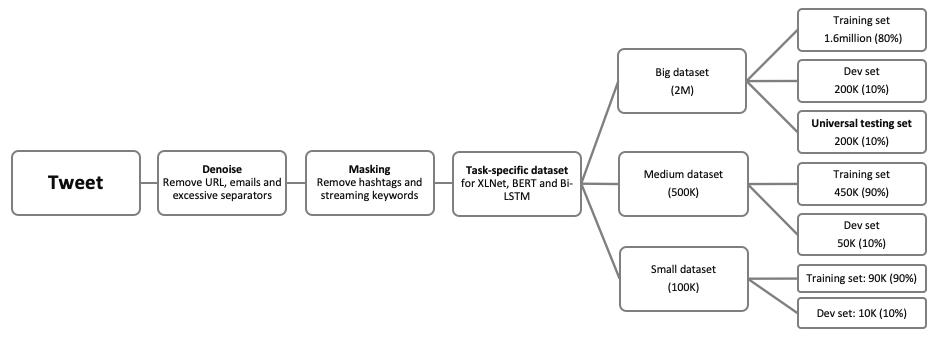}
    \vspace{-0.5cm}
    \caption{Flow chart of data processing methods in this study.}
    \label{fig:data_processing}
\end{figure*}

\subsection{Data Collection and Pre-processing}

The related tweets (Twitter posts) were crawled with the Tweepy stream API using keyword filtering.
Simultaneous streams are collected to build the controversial dataset (CD) and the non-controversial dataset (ND) from March 23 – April 5. The controversial keywords consist of ``Chinese virus” and ``\#ChineseVirus”, whereas non-controversial keywords include ``corona”, ``covid-19”, ``covid19”, ``coronavirus”, ``\#Corona” and ``\#Covid 19”. Only English tweets are collected. We remove any post that contains both controversial keywords and non-controversial keywords in the best effort to separate ``controversial tweets" from non-controversial ones. In total, 2,607,753 tweets for CD and 69,627,062 tweets for ND are collected. We then randomly sample a large dataset (2 million), a medium dataset (500k data), and a small dataset (100k) from the two datasets, respectively. All datasets are perfectly balanced between CD and ND. For preprocessing, URLs, emails, and newlines are removed, as they are not informative for language analysis.

\begin{figure}[htbp]
    \centering
    \includegraphics[width=\linewidth]{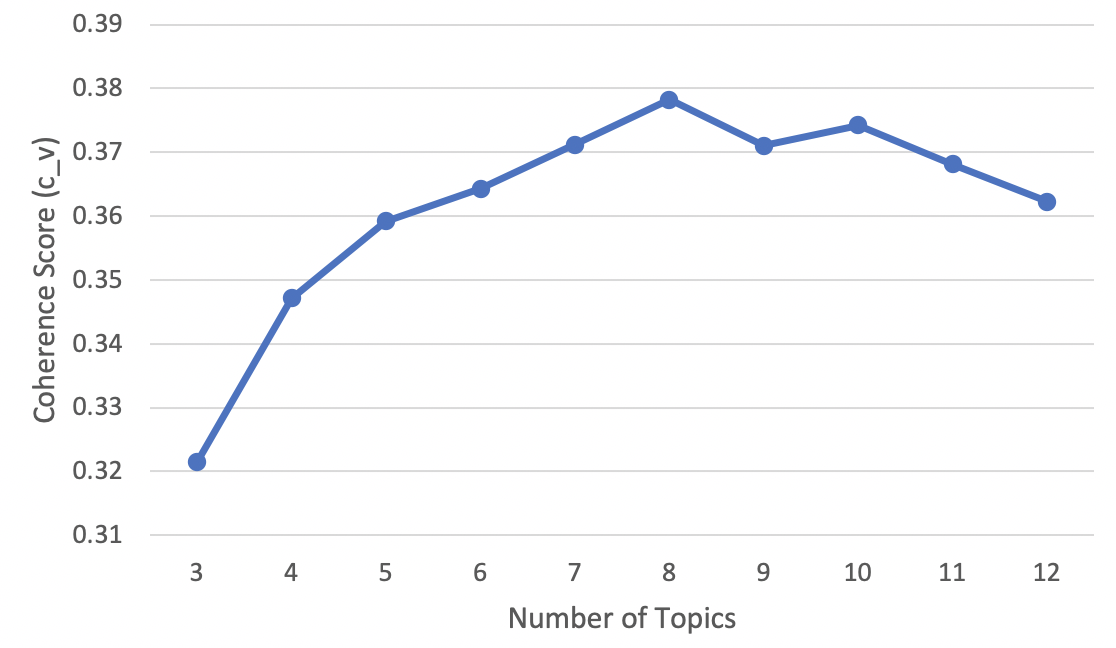}
    \caption{Coherence scores of LDA models with respect to number of topics. In the end, num\_topics=8 is chosen with coherence score \textit{C\textsubscript{v}=0.378}.}
    \label{fig:lda_coherence}
\end{figure}

\subsection{LDA}

We use Latent Dirichlet Allocation to extract topics from the tweets in CD and ND. To see the difference in discussed topics between CD and ND, we merge the CD and ND dataset to generate one LDA model. We utilize our medium dataset (500k) from both CD and ND, resulting in a 1-million dataset. The hyperparameters are tuned with the objective of maximizing the coherence score of C\textsubscript{v}\footnote{C\textsubscript{v} is a performance measure based on a sliding window, one-set segmentation of the top words and an indirect confirmation measure that uses normalized pointwise mutual information (NPMI) and the cosine similarity.}. Coherence scores with respect to a different number of topics are shown in Figure~\ref{fig:lda_coherence}. In the end, we choose \textit{num\_topics=8}, with coherence score \textit{C\textsubscript{v}=0.378}. Since the objective of topic modeling is to find what people talk about when using controversial or non-controversial terms, we also mask all the appearances of the aforementioned streaming keywords by deleting them out of the documents. Next, n-grams are applied on the document. We then perform a comparative analysis to find differences and similarities of topics with documents from the two datasets with topic words weights and t-distributed stochastic neighbor embedding~\citep{maaten2008visualizing} (t-SNE) visualization, an unsupervised machine learning method that visualizes high dimensional data into a low dimensional form.



\subsection{LIWC2015}

Linguistic Inquiry and Word Count (LIWC2015) is applied to extract the sentiment of the tweets of CD and ND. LIWC2015 is a dictionary-based linguistic analysis tool that can count the percentage of words that reflect different emotions, thinking styles, social concerns, and capture people's psychological states\footnote{https://liwc.wpengine.com/how-it-works/}. We focus on 4 summary linguistic variables and 12 more detailed variables that reflect psychological states, cognition, drives, time orientation, and personal concerns of the Twitter users of both groups. We follow the similar methodology used by \citet{yu2008exploring} by concatenating all tweets posted by the users of CD and ND, respectively. One text sample is composed of all the tweets from the aforementioned sampled dataset of CD, and the other is composed of all the tweets from that of ND. We then apply LIWC2015 to analyze these two text samples. In the end, there are 16 linguistic variables for the tweets of both groups. 

\subsection{Classification}

We design a classification task for two purposes. First, we attempt to see if the context of posts can provide sufficient clues to differentiate CD and ND tweets, in the absence of the keywords\footnote{The streaming keywords for data streaming purposes.} in question. Second, we intend to understand whether the state-of-the-art natural language processing tools can be used to classify posts with and without the controversial terms. Here we examine the possible controversial nature of ``Chinese Virus'' from a different angle. Using ``COVID-19'', the official non-controversial term, as our anchor, we test whether ``Chinese Virus'' is equivalent to ``COVID 19'' in its real-world usage in the sense that they are interchangeable. To operationalize this idea, we first mask out all appearances of our streaming keywords for CD and ND. Our idea of masking comes from the original BERT paper where masked language modeling is used to train the BERT model from scratch~\citep{devlin2018bert}. The difference from the BERT pretraining is that rather than masking random tokens or whole words, we mask out the neural terms and controversial terms of COVID-19. We illustrate our masking with the following sample tweets: 

\bigskip
\textit{``The \textbf{Chinese virus}, originated in Wuhan, has killed thousands of people."}

\textit{``My sister tested positive for \textbf{COVID-19} and she is now in quarantine at her home." }

\bigskip
We replace all streaming keywords with a token ``[MASK]", as follow respectively for the two tweets:

\bigskip
\textit{``The \textbf{[MASK]}, originated in Wuhan, has killed thousands of people."}

\textit{``My sister tested positive for \textbf{[MASK]} and she is now in quarantine at her home." }
\bigskip

With properly processed datasets, the state-of-the-art textual classification models, including BERT and XLNet, are fine-trained. We also include the Bi-LSTM model as a baseline. In the following, we provide the details on the models as well as the datasets:

\smallskip

\noindent\textbf{BERT}~\citep{devlin2018bert}, \textbf{B}idirectional \textbf{E}ncoder \textbf{R}epresentations from \textbf{T}ransformers, is a novel transformer-based model pre-trained on unlabeled text, BooksCorpus~\citep{zhu2015aligning} and English Wikipedia, using the masking language modeling task and the next sentence prediction task. We choose the BERT-Base model, with 12 layers, 768 hidden layer size, and 12 heads. The model is then fine-tuned to allow binary classification on our dataset. 


\smallskip

\noindent\textbf{XLNet}~\citep{yang2019xlnet} is a generalized autoregressive pretraining method that utilizes Transformer-XL. The model empirically outperforms BERT in most NLP tasks. We choose XLNet-Base, the smaller version of XLNet with 12 layers, 768 hidden layer size, and 12 heads, to balance model complexity between XLNet and BERT. The model is also fine-tuned for task-specific binary classification. The model is trained for 1 epoch.


\smallskip

\noindent\textbf{Bi-LSTM}, Bidirectional Long Short Term Memory is a typical recurrent neural network suitable for textual classification. The model is developed as a baseline for comparison with BERT and XLNet. We first tokenize the posts in our dataset using the NLTK Tweet Tokenizer~\citep{loper2002nltk}. Next, GloVe~\citep{pennington2014glove} for Twitter is used for word-to-vector purpose with 100-dimensional embeddings. The Bi-LSTM cell has a hidden size of 128 and dropout=0.2. The Bi-LSTM layer is then linked to a fully connected layer and finally passing through a softmax layer. Binary cross-entropy is used as the loss function, along with ADAM as the optimizer. The model is trained for 10 epochs.


\smallskip

\noindent\textbf{Classification Datasets} The process of denoising and creating dataset for the classification task is shown in Figure~\ref{fig:data_processing}. We also attempt to compare the significance of corpus size on classification performance. Therefore, all three sizes of datasets are used for the models. First, we merge CD and ND datasets for all sizes of corpus respectively to form perfectly balanced datasets (50:50 ratio). Next, a number of words and phrases need to be masked to prevent data leakage. Therefore, the aforementioned streaming keywords are removed from CD and ND, respectively. We also remove hashtags from the dataset, as hashtags contain concise but focused meaning and can potentially be the trigger words for classifications. The datasets are then converted into model-compatible formats. A 90-10 split is made on datasets of different sizes to form training and development sets. We make a universal testing set by splitting the big dataset into a 80-10-10 distribution for training, development and testing set, respectively, so that the evaluation metrics are comparable between different datasets.

\section{Analysis Results}

\begin{table*}[htbp]
    \centering
    \begin{tabular}{p{0.05\linewidth} p{0.2\linewidth} p{0.68\linewidth}}
    \hline
     Topic & Topic Title & Topic Keywords\\
    \hline
        1&Trump and Economy&virus, make, trump, chinese, say, bill, pandemic, people, crisis, use, pay, economy, amp, vote, stop, country, video, president, take, day\\
        \hline
        2&Lie and Racism&virus, call, people, stop, pandemic, kill, infect, let, originate, together, pig, chinese, lie, racist, must, take, government, spread, need, help\\
        \hline
        3&Doctor Fight the Virus&virus, chinese, world, fight, doctor, spread, hospital, say, try, trump, amp, call, new, know, still, clear, let, time, save, combat\\
        \hline
        4&Chinese Government and the Virus Outbreak&virus, world, spread, chinese, must, outbreak, much, humanity, government, people, amp, logical, reached\_beije, responsible, know, propaganda, tweet, drag, strip, cover\\
        \hline
        5&Health Workers&help, need, get, thank, amp, test, work, health, well, support, medical, good, worker, speak, state, fight, stay, home, strong, positive\\
        \hline
        6&Stay Home&virus, say, go, come, know, amp, call, get, take, people, time, keep, be, trump, think, see, home, s, make, life\\
        \hline
        7&Test, Cases and Death&virus, case, death, test, people, report, die, new, positive, day, number, say, week, break, world, confirm, country, total, state, last\\
        \hline
        8&Anecdotes and Reports&say, man, send, hospital, right, pass, would, go, chinese, take, handle, medium, do, situation, donate, today, give, agree, month, deadly\\
        \hline
    \end{tabular}
    \caption{Average weight of documents for each topic, normalized. With proportion z-test, significantly more documents from CD discuss about topic 2, 3 and 4 than documents from ND, while significantly more documents from ND discuss about topic 5, 6 and 7 than documents from CD ($p<0.001$). No significant difference discovered for topic 1 and 8.}
    \label{tab:lda_result}
\end{table*}

\subsection{LDA}

\begin{figure}[htbp]
    \centering
    \includegraphics[width=0.8\linewidth]{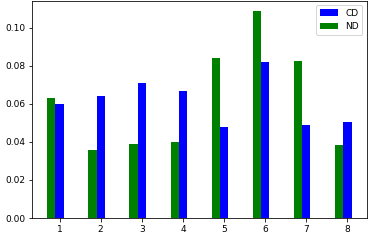}
      \vspace{-0.3cm}
    \caption{Weights of the documents of CD and ND for all topics. By proportion z-test for significance, topic 2, 3 and 4 have significantly more weight of CD tweets, while topic 5, 6 and 7 have more weight of ND tweets. No difference is discovered for topic 1 and 8 between CD and ND.}
    \label{fig:topic_distribution}
\end{figure}

\begin{figure}[htbp]
    \centering
    \includegraphics[width=0.8\linewidth]{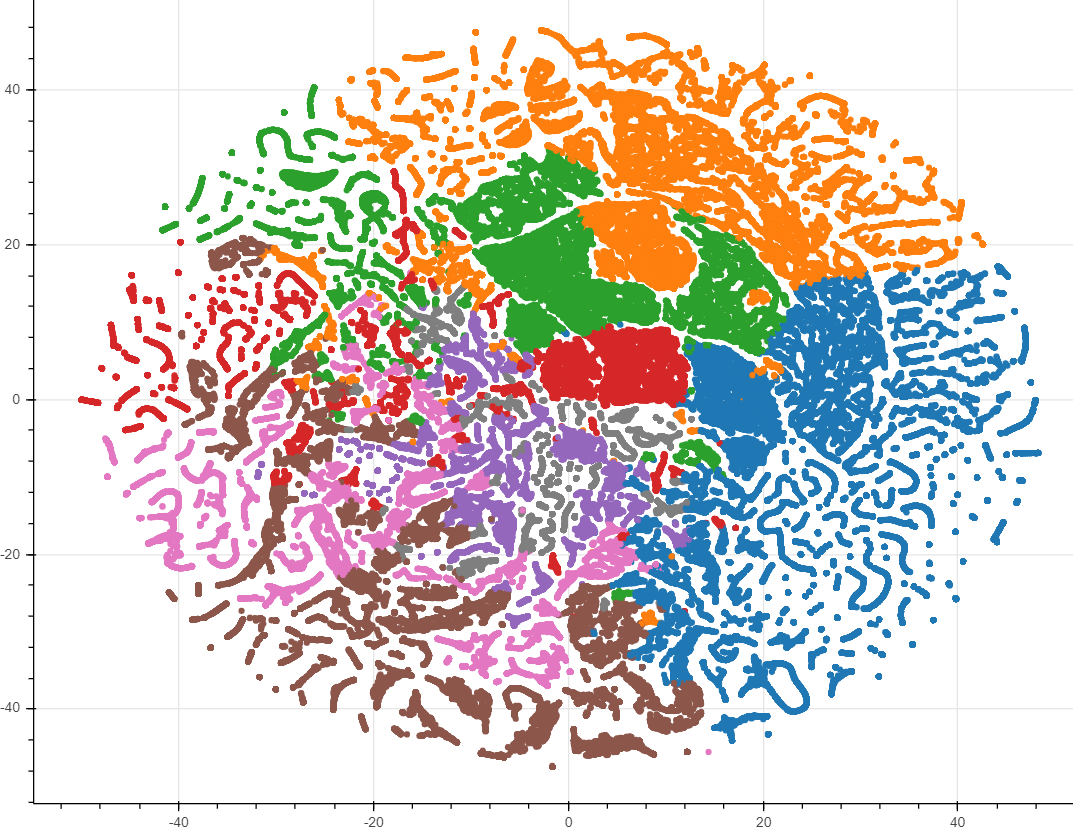}
      \vspace{-0.3cm}
    \caption{t-SNE visualization for LDA. Topic 1-8 are represented by color of blue, orange, green, red, purple, brown, pink and grey, respectively. From t-SNE visualization, topic 2, 3 and 4 are relatively close, while topic 5, 6 and 7 are relatively close. Topic 1 stands out as a big portion, while topic 8 is scattered in the middle of topics. Such finding is incongruous with previous findings from the proportion z-test for significance of CD and ND in each topic.}
    \label{fig:tsne}
\end{figure}

The eight topics generated by the LDA model are reported in Figure~\ref{tab:lda_result}, with the top 20 topic words in each topic. We first discover the distribution of documents associated with each topic by summing and averaging the weight of documents for each topic, as shown in Figure~\ref{fig:topic_distribution}. Proportion z-test is then performed to find significant difference between CD and ND for each topic. We identify that topic 2, 3, and 4 are discussed  significantly more in CD, while topic 5, 6, and 7 are  discussed  more significantly in ND. No significant difference is discovered for topic 1 and 8 between the two datasets. Such findings can be supported by the t-SNE distribution of topics by the LDA model. As shown in Figure~\ref{fig:tsne}, topic 2 (orange), 3 (green) and 4 (red) are more closely discussed together, while topic 5 (purple), 6 (brown) and 7 (pink) are closest to each other. Topic 1 (blue) stands out as a big portion, while topic 8 (grey) is scattered in the center.

Next, we manually assign each topic a topic name to generalize what would most likely be discussed under the topic by looking at topic words that either explicitly refer to a person or an entity that is closely associated with the pandemic, or contain significant emotion or opinion. We then analyze the topics that have significant discussion from only one dataset but not both, then comparing such topics to evaluate different topics in the discussion of CD and ND tweets.

We first consider topics that are CD dominant. Topic 2 (Lie and Racism) is CD dominant and contains very strong opinion words, such as ``lie" and ``racist". It also has ``government", ``let", and ``kill" as keywords, indicating  likely  discussion about how government's decision or behavior "let people killed" by the pandemic. As another CD dominant topic, topic 4 (Chinese Government and the Virus Outbreak) contains more specific keywords, including ``chinese", ``government", ``must" and ``responsible", which indicate how the Chinese government must be held responsible for the spread of the pandemic. It also contains more opinionated keywords such as "propaganda" and ``cover", suggesting misinformation shared by the Chinese government and, to some degree, mis-endeavor by the government to ``cover up" the situation in the early stage of the pandemic. The other CD dominant topic, topic 3 (Doctor Fight the Virus) is the most neutral of the three, with focuses on ``doctor", ``hospital", and ``combat" against COVID-19.

On the other side, the ND dominant topics tend to be more factual. In topic 5 (health workers), most keywords are about doctors and health workers trying to give ``medical", ``support", or ``help" to the patients. In topic 6\footnote{The titling of this topic is rather difficult, as most keywords do not contain significant meaning associated with COVID-19. We finally choose ``Stay Home" as the topic because a keyword ``home" is in the topic.} (stay home), few meaningful keywords associated with COVID-19 are found, except ``Trump" and ``home". In topic 7 (Test, Cases and Death), a large number of keywords are about testing and positive cases (e.g. ``test", ``case", ``report", ``new", ``positive", ``confirm", and ``total"), along with ``death".

One finding is that all three CD dominant topics contain the topic word ``chinese” in the top keywords, even though we have removed all keywords/phrases related to ``Chinese virus" in the documents for LDA. This suggests that the discussions in CD are closely related to China or the Chinese people/government. In addition, two of the three CD dominant topics have keyword ``government", while none in ND dominant topics except one keyword ``Trump" in topic 6. Such difference indicates that the discussion of ND is more factual, while that of CD tends to be more political.

These discrepancies in the topic modeling result contradict the claim of ``only referring to the geo-locational origin of the pandemic" by some public figures who employed the use of ``Chinese virus" when referring to COVID-19.\footnote{https://www.youtube.com/watch?v=E2CYqiJI2pE} Nevertheless, such words have provoked, to a certain degree, racist or xenophobic opinions and hate speeches towards China or people with Chinese ethnicity on social media. Furthermore, hate speeches can spread extremely fast on online social media platforms and can stay online for a long time~\citep{gag2015counter}. \citet{gag2015counter} found that such speeches are also itinerant, meaning that despite forcefully removed by the platforms, one can still find related expression elsewhere on the Internet and even offline.

\subsection{LIWC Sentiment Features}

\begin{figure}[htbp]
    \centering
    \includegraphics[width=0.95\linewidth]{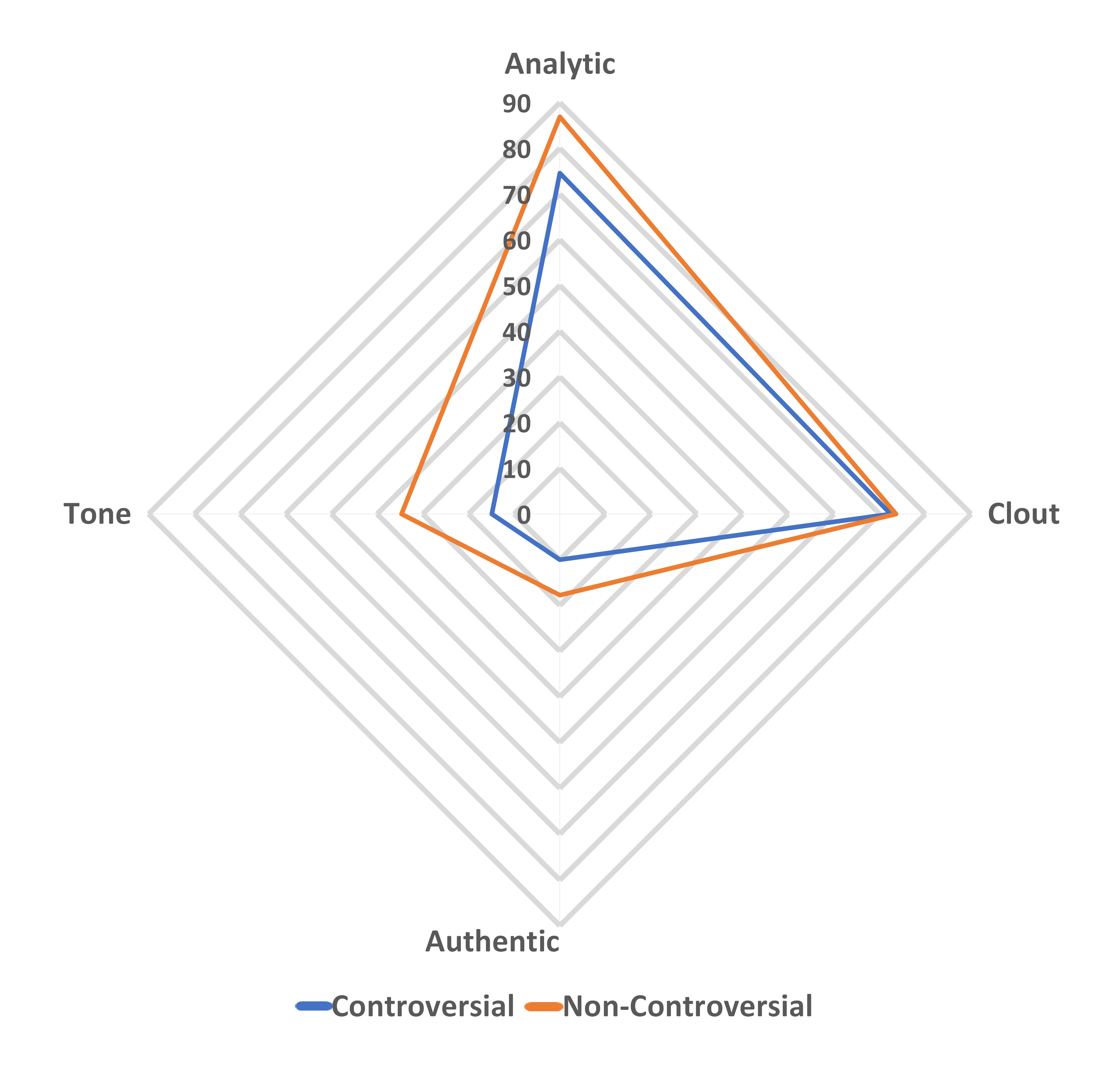}
      \vspace{-0.3cm}
    \caption{Summary linguistic variables for CD/ND. }
    \label{fig:liwc_summary}
   \vspace{-0.3cm}
\end{figure}

Figure~\ref{fig:liwc_summary} shows four summary linguistic variables for CD and ND. We observe that the clout scores for CD and ND are similar. A high clout score suggests that the author is speaking from the perspective of high expertise~\citep{pennebaker2015development}. At the same time, analytical thinking, authentic and emotional tones scores for ND are higher than those for CD. The analytical thinking score reflects the degree of hierarchical thinking. A higher value indicates a more logical and formal thinking~\citep{pennebaker2015development}. A higher authentic score suggests that the content of the text is more honest, personal, and disclosing~\citep{pennebaker2015development}. The emotional tone scores for CD and ND are both lower than 50, indicating that the overall emotions for CD and ND are negative. This is consistent with our expectation. However, the emotional tone score for ND is higher than that for CD, indicating that the Twitter users in ND are expressing relatively more positive emotion.

\begin{table}[htbp]
\setlength{\tabcolsep}{2em}
    \centering
    \begin{tabular}{c c c}
    \hline\hline
    Variables & CD & ND\\
    \hline
        i & 0.96 & 1.04\\
        we &  1.25 & 1.00\\
        she/he & 0.69 & 0.70\\
        they & 1.05 & 0.71\\
        present orientation& 9.37 & 9.22\\
        \hline\hline
    \end{tabular}
    \caption{Scores of ``i", ``we", ``she/he", ``they", and present orientation.}
    \label{tab:five_more}
    \vspace{-0.3cm}
\end{table}

\begin{figure}[htbp]
    \centering
    \includegraphics[width=0.95\linewidth]{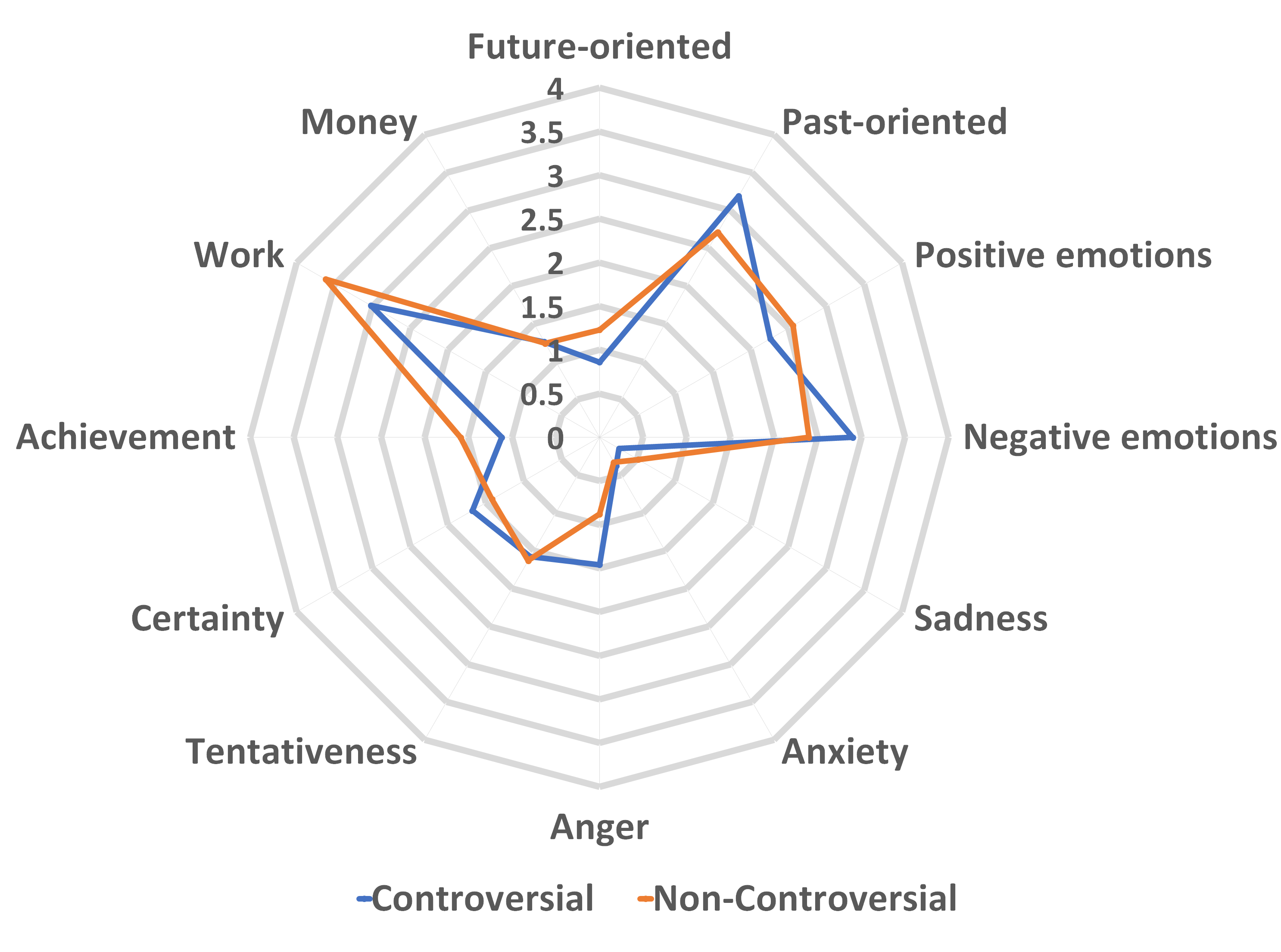}
      \vspace{-0.3cm}
    \caption{Linguistic profiles for the tweets of CD/ND.}
    \label{fig:liwc_multi}
    \vspace{-0.2cm}
\end{figure}

Figure~\ref{fig:liwc_multi} shows 12 more detailed linguistic variables of the tweets of CD and ND. The scores of ``future-oriented" and ``past-oriented" reflect the temporal focus of the attention of the Twitter users by analyzing the verb tense used in the tweets~\citep{tausczik2010psychological}. The tweets of ND are more future-oriented, while those of CD are more past-oriented. To better understand this difference, we conduct a similar analysis as \citet{gunsch2000differential}. We extract 5 more linguistic variables, including four pronoun scores and a one-time orientation score. The scores of ``i", ``we", ``she/he", ``they", and present-orientation are shown in Table~\ref{tab:five_more}. The tweets of CD show more other-references (``they"), whereas more self-references (``i", ``we") are present in the tweets of ND. The scores of ``she/he" of CD and ND are close. The score of present orientation of CD is higher than that of ND. From this observation (similar to the findings of \citet{gunsch2000differential}), we can infer that the tweets of CD focus on the past and present actions of the others, and the tweets of ND focus on the future acts of themselves. Research shows that LIWC can identify the emotion in language use~\citep{tausczik2010psychological}. From the aforementioned discussion, the tweets of both CD and ND are expressing negative emotion, and the emotion expressed by the Twitter users of ND is relatively more positive. This is consistent with the positive emotion score and negative emotion score.

However, there are nuanced differences across the sadness, anxiety, and anger scores. When referring to COVID-19, the tweets of ND express more sadness and anxiety than those of CD do. More anger is expressed through the tweets of CD. The certainty and tentativeness scores reveal the extent to which the event the author is going through may have been established or is still being formed~\citep{tausczik2010psychological}. A higher percentage of words like ``always" or ``never" results in a higher score for certainty, and a higher percentage of words like ``maybe" or ``perhaps" leads to a higher score for tentativeness~\citep{pennebaker2015development}. We observe a higher tentative score and a higher certainty score for the tweets of CD, while these two scores for the tweets of ND are both lower. We have an interesting hypothesis for this subtle difference. Since 1986, \citet{pennebaker2015development} have been collecting text samples from a variety of studies, including blogs, expressive writing, novels, natural speech, New York Times, and Twitter to get a sense of the degree to which language varies across settings. Of all the studies, the tentative and certainty scores for the text of the New York Times are the lowest. However, these two scores for expressive writing, blog, and natural speech are relatively higher. This observation leads to our hypothesis that the tweets of CD are more like blogs, expressive writing, or natural speeches that focus on expressing ideas, whereas the tweets of ND are more like newspaper articles that focus on describing facts.

As for the score of ``achievement", \citet{mcclelland1979inhibited} found that the stories people told in response to drawings of people could provide important clues to their needs for achievement. We hypothesize that the higher value of the ``achievement" score for the tweets of ND reflects the need of these Twitter users to succeed in fighting against COVID-19. As for personal concerns, the scores of ``work" and ``money" of ND are both higher than those of CD, which shows that the Twitter users of ND focus more on the work and money issue (e.g. working from home, unemployment). According to the reports of the U.S. Department of Labor, the advance seasonally adjusted insured unemployment rate was 8.2\% for the week ending April 4. It is interesting to note that the previous high was 7.0\% in May of 1975.\footnote{https://www.dol.gov/ui/data.pdf}


    


\subsection{Classification}

Textual classification for predicting whether a post uses controversial terms associated with COVID-19 is used to test whether neutral terms and their controversial counterparts are linguistically interchangeable. For details on data processing procedures and masking, we refer the readers to the Data and Methodology section. We assume that a low classification accuracy
can be interpreted as easy interchangeability of the two groups of terms. This would support the view that controversial terms such as ``Chinese Virus'' is simply ``COVID-19'' plus its origin. A high accuracy, by contrast, would indicate low interchangeability and suggest that strong differentiating features exist in the context of usage. This would support the argument that ``Chinese Virus" is no substitute for ``COVID-19.''

We report our classification results in Table~\ref{tab:classification}. For robustness, we use three models with different discriminating powers and three different sizes of the dataset. We use F1 score as the metric for model evaluation.

Two immediate observations follow. First, our 1-layered Bi-LSTM model performs relatively poorly for the classification task. BERT-Base performs much better and XLNet is the best. Second, across these three models,  a larger dataset can consistently help achieve a higher F1 score.

Based on our experiments, the highest F1 score we are able to achieve is 0.9521 by XLNet with 500K training samples. Substantively, this high accuracy indicates a low interchangeability between the two groups of terms and supports the view that ``Chinese Virus" is not a straightforward substitute for ``COVID-19.''


\begin{table}[htbp]
    \centering
    \setlength{\tabcolsep}{1em}
    \begin{tabular}{c c c c}
    \hline
     & \textbf{100K} & \textbf{500K} & \textbf{2M}\\
    \hline
        Bi-LSTM	& 0.6723 & 0.6831 & 0.7050\\
        BERT-Base, Cased & 0.8734 & 0.9136 & 0.9302\\
        XLNet-Base, Cased & \textbf{0.9499} & \textbf{0.9521} & -\\

        \hline
    \end{tabular}
    \caption{F1 scores of classification models with different dataset sizes (100K, 500K and 2M). No attempt is made with XLNet on the 2M dataset due to limited computing power. Best results are marked in bold.}
    \label{tab:classification}
\end{table}


\section{Discussion, Conclusion and Future Work}

We have presented a study on the topic preference related to the use of controversial and non-controversial terms associated with COVID-19 on Twitter during the ongoing COVID-19 pandemic. We conclude that, instead of a mere description of the geographical origin of the virus, the use of controversial terms associated with COVID-19 does convey some emotion, and the use of these terms is not the same as the use of non-controversial terms. First, an LDA model is used to extract topics from the controversial and non-controversial posts crawled from Twitter and then qualitatively compare them across the two sets of posts. We find that discussions in the controversial posts are more related to China and, in some cases (2 out of 3), more related to the Chinese government, with mostly criticizing tones, even after the keywords related to ``Chinese virus" are removed before the analysis, whereas discussions in non-controversial posts are more factual and related to fighting the pandemic in the US. Among topics that have significantly more CD posts, one topic contains strong opinionated keywords such as ``lie" and ``racist", while another one contains intensive discussion around the Chinese government, with keywords such as ``government", ``responsible", ``propaganda", and ``cover". In contrast, for topics that have significantly more ND posts, almost all topics are related to discussions on ``facts", such as topic 5 (health workers) and topic 7 (cases and death), with very few opinionated keywords. We then verify that the discrepancies exist between CD and ND posts with a very high performance by our classification models, which analyze the context of the posts in the absence of the controversial and non-controversial terms. 

Furthermore, We find differences across the sentiment of the tweets posted by the users using controversial terms and the users using non-controversial terms. Both groups express negative emotions, yet the tweets of ND are relatively more positive. The tweets of ND also show more analytical thinking and are expressed in a more truthful manner. The tweets of CD focus more on the past and present actions of others, while the tweets of ND focus more on the future acts of the authors themselves. More anger is present in the tweets of CD, while more anxiety and sadness are observed in the tweets of ND. More tentativeness and certainty are observed in the tweets of CD, which is not contradictory since these two scores are both higher in the text samples from blogs and expressive writings that focus on expressing ideas and opinions. These two scores are both lower for the tweets of ND, which is similar to the case of newspaper articles like New York Times. Tweets of ND reflect a strong need for achievement. As for personal concerns, the users of ND focus more on work and money issues.

Our findings can be backed by the recent re-emergence of the use of controversial terms in public. President Trump, in his recent speech\footnote{https://www.cnn.com/2020/08/28/politics/donald-trump-speech-transcript/index.html} at the Republican National Convention, mentioned China for more than 20 times and used the term ``China Virus" multiple times to pass hostile emotions towards China to his supporters. Our study finds a spreading emotion among the public when using the controversial terms, showing how easily such terms can circulate with strong emotions on social media.

Future studies on the uses of controversial terms associated with COVID-19 could investigate any temporal changes in the textual characteristics of social media posts, in response to the development of the COVID-19 pandemic in the United States and other significant events (e.g. changes in wordings from the administration). In addition, other research questions of political and social science can be answered using the collected dataset, such as the connection between nationalism and racism, where the former is prevalent in the current US politics and the latter is present in the use of the controversial terms associated with COVID-19. In addition, since reliable classification performance is achieved with datasets of substantial sizes, future studies could be directed to the application of such classifiers to the monitoring of controversial speeches on social media, with more advanced techniques, such as few-shot learning, to reduce the training cost and adapt to rapid changes in discussions on social media.


\bibstyle{aaai21}
\bibliography{aaai2021}

\appendix

\end{document}